\documentstyle[emulateapj,epsfig]{article}

\begin{document}

\title{Generic Spectrum and Ionization Efficiency of a Heavy Initial 
Mass Function for the First Stars 
}

\author{Volker Bromm$^{1}$}
\affil{Department of Astronomy, Yale University, New Haven, CT 06520-8101;
volker@ast.cam.ac.uk}

\author{Rolf P. Kudritzki}
\affil{
Institute for Astronomy, University of Hawaii, 2680 Woodlawn Drive,
Honolulu, HI 96822;
kud@ifa.hawaii.edu}

\and
\author{Abraham Loeb}
\affil{Astronomy Department, Harvard University, 60 Garden Street, Cambridge, MA 02138;
aloeb@cfa.harvard.edu}

\begin{abstract}
We calculate the generic spectral signature of an early population of
massive stars at high redshifts. For metal-free stars with mass above $
300 M_{\odot}$, we find that the combined spectral
luminosity per unit stellar mass is almost independent of the mass
distribution of these stars.  To zeroth order, the generic spectrum
resembles a black-body with an effective temperature of $\sim 10^5$K,
making these stars highly efficient at ionizing hydrogen and helium.  The
production rate of ionizing radiation per stellar mass by stars more
massive than $\sim 300 M_{\odot}$ is larger by $\sim 1$ order of magnitude
for hydrogen and He I, and by $\sim 2$ orders of magnitude for He II, than
the emission from a standard initial mass function.  This would result in
unusually strong hydrogen and helium recombination lines from the
surrounding interstellar medium. It could also alleviate the current
difficulty of ionizing the intergalactic medium at $z\ga 6$ with the cosmic
star formation rate inferred at somewhat lower redshifts.
\end{abstract}

\footnotetext[1]{Present address: Institute of Astronomy, University of
Cambridge, Madingley Road, Cambridge CB3 0HA, UK}

\section{INTRODUCTION}

One of the most important challenges in modern cosmology is to understand
when and how the cosmic ``dark ages'' ended (Loeb 1998, 1999; Rees 1999). The
absence of a Gunn-Peterson trough in the spectra of high-redshift quasars
implies that the universe was reionized at a redshift $z> 5.8$ (Fan et
al. 2000). An early, pregalactic generation of stars (the so-called
Population III) has long been suspected to control this process (e.g.,
Carr, Bond, \& Arnett 1984; Couchman \& Rees 1986; Haiman \& Loeb
1997). The feedback of the first generation of stars on the ionization
history of the intergalactic medium (IGM) depends crucially on their
initial mass function (IMF). Recent numerical simulations presented
evidence that the primordial IMF might have favored massive stars with a
mass $\ga 10^2 M_\odot$ (Bromm, Coppi, \& Larson 1999, 2001; Abel, Bryan,
\& Norman 2000).  This prediction relies on the existence of a
characteristic scale for the density and temperature of the primordial gas,
and consequently of a characteristic Jeans mass of $M_{J}\sim
10^{3}M_{\odot}$ (see also Larson 1998).  At present, it is not known with
any certainty, what fraction of a Jeans-unstable clump of gas will
eventually be incorporated into the resulting star.  This is due to the
complicated physics of the protostellar feedback on a dust-free envelope,
and the accretion from it. 

Here we adopt the view that the simulated clump masses are indicative of
the final stellar masses, and explore the implications.  Recent
observations of the formation process of present-day stars seem to support
this view.  Motte, Andr\'{e}, \& Neri (1998) have mapped the nearby $\rho$
Ophiuchi star forming cloud in the 1.3 mm dust continuum. Their study
identifies a number of dense, gravitationally-bound clumps with masses
close to stellar values, which appear to lack the emission from embedded
stellar sources. These starless clumps, therefore, might be the direct
progenitors of individual stars.

Ultimately, the question of how massive the first stars were, has to be
answered by observations. It is therefore important to ask whether future
observatories, such as the {\it Next Generation Space Telescope} (NGST),
could place any constraints on the primordial IMF. To address this
question, we have constructed models for the interior structure, and the
emerging spectra of massive Population III stars. Since metal-free stars
have systematically higher effective temperatures ($T_{eff}\sim 10^{5}$ K
for $M > 100 M_{\odot}$), they are very efficient at producing photons
capable of ionizing hydrogen and helium. The reionization of the
intergalactic helium has traditionally been attributed to a population of
quasars. A generation of very massive Population III stars would provide an
important alternative channel, operating at redshifts higher than what is
typically assumed for quasars.

Recently, Tumlinson \& Shull (2000) have discussed the case of metal-free
stars with a normal (Salpeter-like) IMF, up to masses of $90 M_{\odot}$
(see also Ciardi et al. (2000b) for the effect of introducing
a lower mass
cutoff at $M_{c}=5 M_{\odot}$).
It is important, however, to extend this mass range into the regime of very
massive stars with $M > 100 M_{\odot}$, since both the nature of the
feedback on the IGM, as well as the spectral characteristics, change in a
significant way for this regime.

In \S 2, we describe our models for the stellar interior and for the
spectrum.  In \S 3 and \S 4 we
address the ionization efficiency
and discuss the observational signatures, 
respectively.  Finally, \S 5 summarizes
the implications of our results.

\section{Stellar  Models}
\subsection{Stellar Structure}

We construct models for static, metal-free stars in complete (i.e.,
dynamical and thermal) equilibrium by solving the time-independent
equations of stellar structure (e.g., Schwarzschild 1958; Kippenhahn \&
Weigert 1990). The integration is performed in the standard way by fitting
solutions for the core and envelope at a suitably chosen intermediate
point, and the considered mass range is: $100 \leq M/M_{\odot} \leq 1000$.
%REVISED%%%%%%%%%%%%%%%%%
Stars in this mass range have traditionally been termed Very Massive Objects,
or VMOs 
(Bond, Arnett, \& Carr 1984).
%REVISED%%%%%%%%%%%%%%%%%
For such a massive star with the concomitant high interior temperatures,
the only effective source of opacity is electron scattering, given by
$\kappa=0.2 (1 + X)$ cm$^{2}$ g$^{-1}$, where $X\simeq 0.76$ is the mass
fraction in hydrogen.  Other sources of opacity are not included in our
calculations.

In the absence of metals, and in particular of the catalysts necessary for
the operation of the CNO cycle, nuclear burning proceeds in a non-standard
way.  At first, hydrogen burning can only occur via the inefficient $pp$
chain. To provide the necessary luminosity, the star has to reach very high
central temperatures ($T_{c}\simeq 10^{8.1}$ K). These temperatures are
high enough for the simultaneous occurrence of helium burning via the
triple-$\alpha$ process. After a brief initial period of triple-$\alpha$
burning, a trace amount of heavy elements has been formed. Subsequently,
the star follows the CNO cycle. In constructing our main-sequence models,
we therefore assume a level of pre-enrichment with metals, amounting to a
mass fraction of $Z=10^{-9}$.  The resulting models consist of a convective
core, containing 90--$100\%$ of the mass, and a thin radiative
envelope. Due to the high mass and temperature, the stars are dominated by
radiation pressure, and have luminosities close to the Eddington limit:
$L\simeq L_{\mbox{\scriptsize EDD}}=1.25\times 10^{38}$ erg s$^{-1}$
($M/M_{\odot}$).

In this work, we do not include the evolution of the stars away from the
main sequence, which crucially depends on the poorly understood mass-loss
mechanism. While this is clearly an idealization, it is justified by the
fact that the stars spend a major fraction of their total lifetime,
estimated to be $\tau \simeq 0.007 M c^{2}/ L_{\mbox{\scriptsize EDD}}\sim
3\times 10^{6}$ yr, near the location of the main-sequence.

In Figure 1, we show the resulting sequence of models for Population III
stars, and compare to the corresponding Population I sequence (calculated
with $Z=0.02$).  Our results agree well with those of El Eid, Fricke, \&
Ober (1983) in the mass range $100 < M/M_{\odot} < 500$.  It is evident
that the respective effective temperatures approach an asymptotic value
which is independent of mass, $T_{eff}\simeq 1.1\times 10^5$ K for
Population III, and $T_{eff}\simeq 6.5\times 10^4$ K for Population I. This
characteristic behavior of very massive stars derives from the assumptions
of hydrostatic equilibrium, radiation pressure support with a luminosity
close to the Eddington limit, and the nuclear physics of hydrogen burning.
%REVISED%%%%%%%%%%%%%%%%%%%
The peculiar behavior of very massive Population III stars has already
been discussed by a number of previous authors (see
Bond, Arnett, \& Carr (1984) and references therein).
Bond, Arnett, \& Carr (1984), in particular, have pointed out that the
effective temperature, $T_{eff}\sim 10^{5}$ K, is almost independent of
mass for $M > 100 M_{\odot}$.
For completeness, we here give a somewhat simplified version of the argument.
%REVISED%%%%%%%%%%%%%%%%%%%

\placefigure{fig1}

The mechanical structure of a massive, radiation-pressure dominated star is
approximately given by a polytrope of index $n=3$ (see Kippenhahn \&
Weigert 1990). If $w(z)$ is the corresponding solution of the Lane-Emden
equation, and $z=(z_{3}/R) r$ the dimensionless radial coordinate, one can
express the density and pressure as $\rho=\rho_{c}w^{3}$, and $P=(\pi G
R^{2} \rho_{c}^{2/3}/z_{3}^{2}) \rho^{4/3}$, respectively.  Here,
$z_{3}=6.89685$ is the dimensionless radius of the polytrope,
$\rho_{c}=54.1825(3 M/4\pi R^{3})$ the central density, and the other
symbols have their usual meaning. With the polytropic value for the central
pressure, $P_{c}\simeq 11.1 G M^{2}/R^{4}$, one can estimate the
temperature at the center of the star, $T_{c}=(3P_{c}/a)^{1/4}$, to be
\begin{equation}
T_{c}\simeq 8.4\times 10^{7}\mbox{\, K\,}
\left(\frac{M}{100 M_{\odot}}\right)^{1/2}
\left(\frac{R}{10 R_{\odot}}\right)^{-1}
\mbox{\ \ \ .}
\end{equation}
To further constrain $T_{c}$, we follow Fowler \& Hoyle (1964), and
consider the global energy balance of the star.  The average nuclear energy
generation rate (in erg s$^{-1}$ g$^{-1}$) is
$\bar{\epsilon}=L/M=\int_{0}^{M}\epsilon\mbox{d}m$, and we assume
$\epsilon\simeq \epsilon_{\mbox{\scriptsize CNO}}$. Writing
$\epsilon=\epsilon_{c}(\rho/\rho_{c})(T/T_{c})^{\nu}$, with $c$ denoting
central values, one finds
\begin{equation}
\bar{\epsilon}\simeq 162.5 \epsilon_{c}
\left(
\frac{1}{z_{3}^{3}}\int_{0}^{z_{3}}w^{6 + \nu}z^{2}\mbox{d}z
\right)
\mbox{\ \ \ .}
\end{equation}
A reasonable fit to the CNO cycle energy generation rate close to
$\mbox{log} T\simeq 8.1$ is given by
\begin{equation}
\epsilon\simeq 1.3\times 10^{12} Z X \rho
\left(
\frac{T}{10^{8}\mbox{\,K}}
\right)^{8}
\mbox{\ \ \ .}
\end{equation}
%REVISED%%%%%%%%%%
This form derives from a linear fit to the exact expression for
$\epsilon_{\mbox{\scriptsize CNO}}$ (Kippenhahn \& Weigert 1990,
p. 165), at 
$\mbox{log} T\simeq 8.1$.
%REVISED%%%%%%%%%%
Using $\nu=8$ in equation (2) gives $\bar{\epsilon}\simeq 0.05 \epsilon_{c}$.
From $L\simeq L_{\mbox{\scriptsize EDD}} = \bar{\epsilon}M$ follows the
desired relation
\begin{equation}
T_{c}\simeq 1.9\times 10^{8}\mbox{\, K\,}
\left(\frac{Z}{10^{-9}}\right)^{-1/8}
\left(\frac{M}{100 M_{\odot}}\right)^{-1/8}
\left(\frac{R}{10 R_{\odot}}\right)^{3/8}
\mbox{\ \ \ .}
\end{equation}
Combining equations (1) and (4) gives the mass-radius relation for very 
massive stars
\begin{equation}
M\simeq 370 M_{\odot}
\left(\frac{Z}{10^{-9}}\right)^{-0.2}
\left(\frac{R}{10 R_{\odot}}\right)^{2.2}
\mbox{\ \ \ ,}
\end{equation}
from which it can be seen that approximately $M\propto R^{2}$.
Finally, the effective temperature can be estimated from
$L\simeq L_{\mbox{\scriptsize EDD}}=4\pi R^{2} \sigma T_{eff}^{4}$. Using
equation (5) to eliminate the radius, we find
\begin{equation}
T_{eff}\simeq 1.1 \times 10^5 \mbox{\, K\,}
\left(\frac{Z}{10^{-9}}\right)^{-0.05}
\left(\frac{M}{100 M_{\odot}}\right)^{0.025}
\mbox{\ \ \ .}
\end{equation}
This analytical estimate for $T_{eff}$, and the very weak dependence on mass,
are in good agreement with the numerical results in Figure 1.

\subsection{Emergent Spectrum}

Having determined the basic stellar parameters,
we now construct model
atmospheres and emergent spectra. The extreme luminosities very close to
the Eddington limit and the very high effective temperatures require the
sophisticated effort of using NLTE unified model atmospheres with spherical
extension and stellar winds, as originally introduced by Gabler et
al. (1989). This new type of model atmospheres provides a smooth
transition from the subsonic, quasi-hydrostatic photosphere to the
supersonic, radiation-driven stellar wind, and allows the simultaneous
treatment of
spectral features formed deep in the photosphere (such as optical
continua and line wings) and those arising from the outer layers,
where the influence of winds becomes significant (such as strong absorption
edges, far infra-red continua and line cores). A detailed discussion is
given by Kudritzki (1998).

For our work, we use the new unified model code originally developed by
Santolaya-Rey, Puls, \& Herrero (1997), and further improved by J. Puls (private
communication).
Bound-bound, bound-free and free-free
absorption by hydrogen and helium, as well as Thomson scattering are
included as opacity sources.  A value of 0.1 has been adopted for the
number ratio of helium to hydrogen.  Changing this ratio towards slightly
smaller values closer to estimates for the primordial helium abundance
would not lead to a significant modification of the calculated spectra.

The atmospheric models are fully specified by the effective temperature,
the gravity and the stellar radius (all taken at the Rosseland optical
depth $\tau_{\rm ross}=2/3$) together with the mass-loss rate $\dot{M}$,
the terminal velocity $v_{\infty}$, and the parameter $\beta$ which
describes the radial slope of the velocity field via
\begin{equation}
v(r) = v_{\infty}(1-b/r)^{\beta}
\mbox{\ \ \ .}
\end{equation}
The constant $b$ is chosen to guarantee a smooth transition into the
hydrostatic stratification at a prespecified outflow velocity smaller than
the isothermal sound speed (normally 0.1 $c_{s}$). Velocity fields
of this form are predicted by the theory of radiation driven winds (e.g.,
Castor, Abbott, \& Klein 1975; Pauldrach, Puls, \& Kudritzki 1986)
and usually lead to an excellent
fit to observed stellar wind lines based on radiative transfer and model
atmosphere calculations (see Kudritzki \& Puls 2000).

At zero metallicity, we expect stellar winds to be of only
minor importance.
However,
a radiatively driven wind might still arise due to the opacity in the He II
resonance lines, in combination with the contribution from bound-free and
free-free absorption. No wind models have been
constructed so far for this extreme case of zero metallicity. Kudritzki
(2000) has developed a new algorithm, which allows to deal with line
driven winds at very low metallicity down to $10^{-4}$ of the solar
value. As a rough extrapolation from his results we adopt $\dot{M} =
10^{-10}M_{\odot}$ yr$^{-1}$, $v_{\infty} = 1900$ km s$^{-1}$ and $\beta =
1$. This choice of stellar wind parameters results in an almost
hydrostatic, but still spherically extended density structure. From a
number of additional test calculations carried out with different wind
parameters we note that the resulting emergent spectra show only a very
small parameter dependence as long as the winds are weak.  For instance,
enhancing the mass-loss rate to a value of $\dot{M} = 10^{-8}M_{\odot}$ has
little influence on the energy distribution and line profiles.

In Figure 2, we present the resulting spectra for different masses. Shown
is the specific flux, $F_{\nu}$, at the surface of the star. At these high
effective temperatures, there is still a rich spectrum of He II and H lines
in the EUV, UV, optical, and IR. Some of these lines show emission features
as a result of an overpopulation of their upper levels due
to NLTE effects.
%%%%REVISED%%%%%%%%%%%%%
The He II resonance lines have emission wings and central absorption cores,
whereas some of the He II Pickering lines have a central emission peak
on top of absorption wings. Such types of line profiles reflect the 
stratification of the ratio of the NLTE departure coefficients of the upper
to the lower line level and are common for very hot stars. A first
interpretation of the underlying physics has been given in the pioneering
paper by Auer \& Mihalas (1972).
%%%%REVISED%%%%%%%%%%%%%
Strikingly, the He II absorption edge at $228
\mbox{\,\AA}$ is very weak or turns into emission for higher masses. This is
again a distinct NLTE effect which has been discussed in detail by Husfeld
et al. (1984) and Gabler et al. (1989, 1992).

\placefigure{fig2}
\placefigure{fig3}

For $M > 300 M_{\odot}$, the spectra become increasingly similar. This 
similarity is even more evident when the resulting specific luminosity, 
$L_{\nu}=4\pi R^{2} F_{\nu}$, is plotted per unit stellar mass, as shown in 
Figure 3. Stars with luminosities close to the Eddington limit, 
$L_{\nu}\propto L_{\mbox{\scriptsize EDD}}\propto M$, approach a universal 
value for their flux per unit mass. The prediction that the total spectral 
luminosity  depends only on the total mass of stars but is {\it independent
of the initial mass function}, is a unique feature of a population of very 
massive stars. In \S 4, we explore the observational 
consequences of this result, but discuss first the production of ionizing
photons from these stars.

\section{FEEDBACK EFFECTS}

The spectrum for a population of very massive, zero-metal stars
deviates most strongly from the case with a Salpeter IMF at short
wavelengths, near the peak of the corresponding black body at
$\lambda_{em}\simeq 250 \mbox{\,\AA}$. An important difference is
therefore expected in the rate of producing ionizing photons.
To quantify this effect for the species H I, He I, and He II, we
calculate the number of ionizing photons per unit stellar mass
according to 
\begin{equation}
\dot{N}_{ion}=\frac{1}{M}\int_{\nu_{th}}^{\infty} \frac{L_{\nu}
\mbox{d}\nu}{h\nu}
\mbox{\ \ \ ,}
\end{equation}
where $\nu_{th}$ is the respective threshold for ionization.
%REVISED%%%%%%%%%%%%
We here make the somewhat idealized assumption that {\it all} Population III
stars were more massive than $300 M_{\odot}$.
Therefore, as demonstrated in the previous section, their spectra, when
scaled to stellar mass, have a generic form which is almost independent
of mass. Specifically, we take the luminosity per unit mass, $L_{\nu}/M$, for a
$1000 M_{\odot}$ star, as given by the dot-dashed line in Fig. 3, to
be representative for the generic form, and use it in evaluating the integral
in equation (8). 
%REVISED%%%%%%%%%%%%
The resulting production rates are presented in Table 1.
Comparing these rates to the corresponding ones for a Salpeter IMF,
as given by Tumlinson \& Shull (2000), we find 
an enhancement in the number of ionizing photons
by a factor of $10-20$ for H I and He I, and by a factor of $\sim 75$ for
He II.
%REVISED%%%%%%%%%%%%
 We have verified that using $L_{\nu}/M$ for the 
$300 M_{\odot}$ case (dotted line in Fig. 3) results in ionization
efficiencies that are at most a factor of two smaller: $1.2\times 10^{48}$
photons s$^{-1}$ $M_{\odot}^{-1}$ for the ionization of H I,
$8.6\times 10^{47}$
photons s$^{-1}$ $M_{\odot}^{-1}$ for He I, and
$2.0\times 10^{47}$
photons s$^{-1}$ $M_{\odot}^{-1}$ for He II.
By comparing to Table 1, one can see that our
choice of the $1000 M_{\odot}$ spectrum as being representative for the mass 
range $1000 M_{\odot} > M > 300 M_{\odot}$
is well justified.
%REVISED%%%%%%%%%%%%

\placetable{tab1}

To illustrate the implications of this result, consider the following
zeroth-order estimate for the fraction of baryons that has to be
incorporated into stars, $f_\star$, in order to reionize the respective
species.  
Assuming that $\eta$ ionizing photons are needed per atom ($\eta>1$ to
account for the effect of recombinations), we find for hydrogen
\begin{equation}
f_{\star}(\mbox{H I})\simeq\frac{\eta X}{m_{\mbox{\scriptsize H}}\tau_
{\mbox{\scriptsize MS}}\dot{N}_{ion}(\mbox{H I})}\simeq 10^{-4} \eta_{10}
\mbox{\ \ \ ,}
\end{equation}
where $\eta_{10}=(\eta/10)$, and $\tau_{\mbox{\scriptsize MS}}\simeq
2\times 10^{6}$ yr is the approximate main-sequence lifetime of a very
massive star. For helium, one has $f_{\star}(\mbox{He I})\simeq
10^{-5}\eta_{10}$ and $f_{\star}(\mbox{He II})\simeq 4\times
10^{-5}\eta_{10}$. The corresponding fractions for a Salpeter IMF are
$f_{\star}\sim 10^{-3}\eta_{10}$.

Consequently, if the first stars are indeed characterized by a heavy IMF,
the intriguing possibility arises that the rare star forming clouds, which
originate from the high-$\sigma$ peaks in the primordial density field,
might have played an important role in the reionization of both hydrogen
and helium at high redshifts. Furthermore, helium might have even been
reionized before hydrogen. Such a sequence of events would be in marked
contrast to the standard scenarios of reionization (e.g., Gnedin \&
Ostriker 1997; Haiman \& Loeb 1997; Gnedin 1998; Ciardi et al. 2000a;
Haiman, Abel, \& Rees 2000; Miralda-Escud\'{e}, Haehnelt, \& Rees
2000). Recently, Madau, Haardt, \& Rees (1999) have pointed out that the
known populations of quasars and star-forming galaxies are not sufficient
to account for the required number of ionizing photons at redshifts $z\sim
5$. An early generation of very massive stars could alleviate this problem,
and contribute a significant part of this unexplained deficit.

\section{OBSERVATIONAL SIGNATURE}

\subsection{Stellar Spectrum}

Since the first clusters of stars form prior to the epoch of reionization,
they are embedded within a neutral IGM. Consequently, the observed flux
shortward of the Ly$\alpha$ resonance wavelength, $\lambda_\alpha=1216$\,\AA,
will be suppressed by the neutral IGM before reionization (Gunn \& Peterson
1965) and by the Ly$\alpha$ forest after reionization. The suppression will
be complete (due to the neutral IGM) for observed wavelengths between
$(1+z_{\rm reion})\lambda_c$ and $(1+z_s)\lambda_\alpha$ and partial (due
to the Ly$\alpha$ forest) at shorter wavelengths (Haiman \& Loeb 1999).
Here, $\lambda_c=912$\,\AA\,is the Lyman-limit wavelength, $z_s$ is the source
redshift, and $z_{\rm reion}$ is the reionization redshift.  The sharp
trough of the observed flux at $(1+z_s)\lambda_\alpha$ can be used to
unambiguously determine the redshift of the source, even in the absence of
strong spectral lines.  To predict the observed spectrum at wavelengths
longer than $\lambda_{\alpha}$, we consider
$L(\nu_{em})$, the monochromatic luminosity per unit stellar mass at the
emitted frequency $\nu_{em}$.
%REVISED%%%%%%%%%%%%%%%%%%%
We again make the assumption that the spectrum for a star of $1000 M_{\odot}$
is representative for the mass range $M > 300 M_{\odot}$.
With the flux at the surface of the $1000 M_{\odot}$ star being $F(\nu_{em})$,
as shown in Figure 2, one has $L(\nu_{em})=4\pi R^{2} F(\nu_{em})$, where
$R\simeq 13.7 R_{\odot}$ is the stellar radius.
%REVISED%%%%%%%%%%%%%%%%%%%
The observed flux, unattenuated by the neutral IGM, at a
frequency $\nu_{obs}=\nu_{em}/(1+z_{s})$ is
\begin{equation}
f(\nu_{obs})=\frac{L(\nu_{em})}{4\pi r^{2}(1+z_{s})}
\mbox{\ \ \ ,}
\end{equation}
where
\begin{equation}
r=\frac{c}{H_{0}}\int_{0}^{z_{s}}\frac{\mbox{d}z}
{\sqrt{\Omega_{m}(1+z)^{3}+\Omega_{\Lambda}}}
\mbox{\ \ \ .}
\end{equation}

In Figure 4, we show the predicted spectrum for a cluster at $z_{s}=10$,
assuming a flat universe with $\Omega_{\Lambda}=1-\Omega_{m}=0.7$ and a
Hubble constant of $H_{0}=65$ km s$^{-1}$ Mpc$^{-1}$.
%%REVISED%%%
We have taken into account the red damping wing of the Gunn-Peterson
trough, using the analytical expression given by Miralda-Escud\'{e} (1998).
In evaluating the optical depth in the vicinity of the Ly$\alpha$ resonance,
we assume $z_{\rm reion}=7$ and $\Omega_{b} h=0.03$, corresponding to
a Gunn-Peterson optical depth of $\tau_{0}(z_{s})=7.4\times 10^{5}$.
We find that the resulting line shape is not very sensitive to the choice of
$z_{\rm reion}$.
%%REVISED%%%

The peak flux is
conveniently located in the spectral range between $1-5\mu$m, where NGST is
expected to reach a photometric sensitivity of $\sim 0.1-1$ nJy.  We plot
the flux per $10^{6}M_{\odot}$, but it is straightforward to adjust the
spectrum to other cases due to the linear scaling with total stellar mass.
The composite spectrum of a Salpeter IMF with stellar masses in the range
$0.1\la M/M_{\odot} \la 100$, has been worked out by Tumlinson \& Shull
(2000), and is also shown in Figure 4, normalized to the same total stellar
mass. In comparing the two spectra, it is evident that the observed flux at
any given wavelength is a factor of $5-10$ larger for the case calculated
with a heavy IMF. The difference in absolute flux, however, will not allow
to discriminate between the two IMFs, since the total stellar mass of the
observed system is not known a priori. One instead has to rely on the
difference in colors.

\placefigure{fig4}

For this purpose, we define the ratios $F_{12}/F_{23}$ and $F_{23}/F_{45}$,
where
$F_{12}$, $F_{23}$, and $F_{45}$ are the integrated fluxes between $1-2\mu$m,
$2-3\mu$m, and
$4-5\mu$m, respectively. The two colors can then be expressed as a difference
in magnitude
\begin{equation}
\Delta m_{A}=-2.5\log\frac{F_{12}}{F_{23}}
\mbox{\ \ \ ,}
\end{equation}
and
\begin{equation}
\Delta m_{B}=-2.5\log\frac{F_{23}}{F_{45}}
\mbox{\ \ \ .}
\end{equation}
Using the spectra in Figure 4, we find these ratios to be 
$F_{12}/F_{23}\simeq 2.7$ and $F_{23}/F_{45}\simeq 9.3$
in the case of a heavy IMF, and
$F_{12}/F_{23}\simeq 2.5$ and $F_{23}/F_{45}\simeq 7.3$
in the case of a Salpeter IMF.
In terms of magnitudes, one has
\begin{displaymath}
\Delta m_{A}\simeq\left\{
\begin{array}{ll}
-1.1
& \mbox{for a heavy IMF } \\
-1.0 &  \mbox{for a Salpeter IMF } \\
\end{array}
\right. 
\mbox{\ \ \ ,}
\end{displaymath}
and
\begin{displaymath}
\Delta m_{B}\simeq\left\{
\begin{array}{ll}
-2.4
& \mbox{for a heavy IMF } \\
-2.1 &  \mbox{for a Salpeter IMF } \\
\end{array}
\right. 
\mbox{\ \ \ .}
\end{displaymath}
Hence, the observed spectrum in the case of a heavy IMF is significantly
bluer.  Redenning by dust is expected to be weak for the first star
clusters (and is already small in some known high-redshift galaxies, see
Kudritzki et al. 2000).
Moreover, the two colors differentiate between the IMFs in a way that is
not degenerate with conventional dust.
To see this, we ask: Can the presence of dust in the cluster with a heavy
IMF cause the two colors to simultaneously agree with those predicted in the
Salpeter case? Applying the average extinction curve for the interstellar
medium of the Milky Way (e.g., Savage \& Mathis 1979), we adjust the 
amount of dust present in the cluster such that color $A$ is reddened to
agree with the prediction for the Salpeter IMF. The corresponding
change in color $B$ results in $\Delta m_{B}\simeq -1.5$, compared
to $\Delta m_{B}\simeq -2.1$ for the Salpeter IMF. Unless one invokes
rather contrived properties for the dust, the photometric signature
of a heavy IMF therefore cannot masquerade as the one expected from a
Salpeter IMF. 
By measuring these colors, NGST
might thus be able to either confirm or exclude the dominance of very
massive stars in the early universe.

\subsection{Strong Emission of Recombination Lines}

The hard UV emission from a star cluster with a heavy IMF is likely to be
reprocessed by the surrounding interstellar medium and produce very strong
recombination lines of hydrogen and helium.
With $\dot{N}_{ion}$ as derived in \S 3,
the emitted luminosity
$L_{line}^{em}$ per unit stellar mass
in a particular recombination line can be estimated as
follows
\begin{equation}
L_{line}^{em} = p_{line}^{em} h\nu \dot{N}_{ion} (1 - p^{esc}_{cont})
p^{esc}_{line}
\mbox{\ \ \ ,}
\end{equation}
where $p_{line}^{em}$ is the probability that a recombination would lead to
the emission of a photon in the corresponding line, $\nu$ is the frequency
of the line and $p^{esc}_{cont}$ and $p^{esc}_{line}$ are the escape
probabilities for the ionizing photons and the line photons,
respectively. It is natural to assume that the stellar cluster is
surrounded by an ionization bounded H II region, 
and hence $p^{esc}_{cont}$ is close to zero 
(Wood \& Loeb 2000; Ricotti \& Shull 2000).  In addition,
$p^{esc}_{line}$ is likely to be close to unity in the H II region, due to
the lack of dust in the ambient metal-free gas. Although the emitted line
photons may be scattered by neutral gas, they will diffuse out to the
observer and eventually survive if the gas is dust free. Thus, for
simplicity, we adopt a value of unity for $p^{esc}_{line}$.

To be concrete, let us consider case B recombination which yields
$p_{line}^{em}$ of about 0.65 and 0.47 for the Ly${\alpha}$ and He II 1640\,\AA
\,lines, respectively. In arriving at these numbers,
we assume an electron temperature of $\sim 3\times
10^4$K,
and an electron density of $\sim 10^{2}-10^{3}$ cm$^{-3}$ inside the H II
region (see Storey \& Hummer 1995). With $\dot{N}_{ion}$ from Table 1, we
then obtain $L_{line}^{em} = 1.7\times 10^{37}$ and $2.2\times 10^{36}$
erg s$^{-1}M_{\odot}^{-1}$ for the recombination luminosity of Ly$\alpha$ and
He II 1640\,\AA\,per embedded stellar mass. A cluster of $10^{6} M_{\odot}$ in
stars would then produce 4.4 and 0.6 $\times 10^{9}L_{\odot}$ in the
Ly$\alpha$ and He II 1640\,\AA\,lines. Similarly high luminosities would be
produced in other recombination lines at longer wavelengths, such as He II
4686\,\AA\,and H$\alpha$.

The rest--frame equivalent width of these emission lines measured against
the stellar continuum of the embedded cluster at the line wavelengths is
given by
\begin{equation}
W_{\lambda} =\left(\frac{L_{line}^{em}}{L_{\lambda}}\right)
\mbox{\ \ \ ,}
\end{equation}
where $L_{\lambda}$ is the spectral luminosity per unit wavelength of the
stars at the line resonance. From Figure 3, we obtain a spectral luminosity
per unit frequency $L_{\nu} = 2.7\times 10^{21}$ and $1.8\times 10^{21}$
erg s$^{-1}$ Hz$^{-1}M_{\odot}^{-1}$ at
the corresponding wavelengths.  Converting to
$L_{\lambda}$, we obtain rest-frame equivalent widths of $W_{\lambda}$ =
3100\,\AA\,and 1100\,\AA\,for Ly$\alpha$ and He II 1640\,\AA\,, respectively. These
extreme emission equivalent widths are more than an order of magnitude
larger than the expectation for
a normal cluster of hot stars with the same total mass and a
Salpeter IMF under the same assumptions concerning
the escape probabilities and recombination (e.g., Kudritzki et al. 2000).
The equivalent widths are, of course, larger by a factor of $(1+z_{s})$ in the
observer frame. Extremely strong recombination lines, such as Ly$\alpha$
and He II 1640\,\AA, are therefore expected to be an additional
spectral signature that is unique to very massive stars in the early
universe.
%REVISED%%%%%%%%%%%%%%%%%
Tumlinson \& Shull (2000) have discussed the possible observational role of
the He II 1640\,\AA\, line in the context of a standard Salpeter IMF.
%REVISED%%%%%%%%%%%%%%%%%

The He II 1640\,\AA\, line would reach the observer unaffected by the intervening
IGM, since its wavelength is longer than that of the Ly$\alpha$ transition
which dominates the IGM opacity. However, the Ly$\alpha$ line emitted by
the source would be resonantly scattered by the intergalactic H I. At
redshifts lower than the reionization redshift, most of the IGM is ionized
and only the blue wing of the line would be suppressed by the Ly$\alpha$
forest. But if the star cluster lies beyond the reionization redshift,
the intervening neutral IGM acts as fog and obscures the view of the source
at the Ly$\alpha$ wavelength. The line photons emitted by the source
scatter until they eventually redshift out of resonance and escape due to
the Hubble expansion of the surrounding intergalactic H I. As a result, the
source creates a faint Ly$\alpha$ halo on the sky (Loeb \& Rybicki 1999).
For a source at $z_{s}\sim 10$, the halo occupies an angular radius of $\sim
15^{\prime\prime}$ and corresponds to a line that is broadened and
redshifted by $\sim 10^3~{\rm km~s^{-1}}$ relative to the source. The
scattered photons are also highly polarized (Rybicki \& Loeb 1999). A heavy
IMF would make the detection of diffuse Ly$\alpha$ halos around the first
star clusters more accessible to future telescopes, and would provide an
important tool for probing the neutral IGM prior to the epoch of
reionization.  Observations which focus directly on the bright central
source are more straightforward, and should show a strong suppression of
the Ly$\alpha$ line by the damping-wing of the ``Gunn-Peterson trough'',
caused by the IGM scattering (Miralda-Escud\'{e} 1998). The very brightest
sources might be an exception, in that the H II region they generate in the
surrounding IGM might shift the Ly$\alpha$ damping wing and allow a portion
of the red wing of the emission line to be observed (Cen \& Haiman 2000;
Madau \& Rees 2000).

\section{SUMMARY AND CONCLUSIONS}

We have explored the consequences of a possible early generation of very
massive, zero-metal stars. These stars are almost fully convective, have
luminosities close to the Eddington limit, and effective temperatures,
$T_{eff}\sim 10^{5}$K, which are almost independent of stellar mass. Using
a spherically extended, NLTE stellar atmosphere code, we have subsequently
constructed the emerging spectra. If normalized by the mass of the star,
the spectrum has a generic form, almost independent of mass for $M\ga
300M_{\odot}$. Based on the photometric colors of the composite spectrum
from a cluster of Population III stars, we find that it would be feasible
observationally to constrain the primordial IMF with future infrared
telescopes such as NGST (Fig. 4). 

We also investigated the ionizing photon production from these massive
stars. Compared to a population of stars with a normal, Salpeter-like IMF,
the rate of producing ionizing photons is enhanced by an order of magnitude
for H and He I, and by a factor of $\sim 100$ for He II. As a result, the
interstellar medium of a star cluster with a heavy IMF is likely to
generate extremely strong recombination lines, such as Ly$\alpha$ and
He II 1640\,\AA\,, with equivalent widths that are larger by at least an order of
magnitude compared to a standard IMF.

Very massive Population III stars might have played a significant role in
the reionization of the hydrogen and helium in the IGM at high redshifts.
The ability to efficiently ionize He II provides an alternative channel for
a process which has otherwise to rely on the presence of quasars with their
non-thermal spectra. Equally important is the possible contribution of
massive stars to the ionizing photon budget at $z\sim 5$, where the known
quasars and galaxies are unable to maintain the IGM in an ionized state as
long as the IMF is assumed to have the standard form (Madau et al. 1999).

The massive stars considered in this work are likely to end up as black
holes (Fryer, Woosley, \& Heger 2000). If $\gamma$-ray bursts (GRBs)
originate from the collapse of these massive stars, then a heavy IMF will
produce a larger abundance of GRBs at high redshifts than previously
estimated (Ciardi \& Loeb 2000). The bright afterglows of these GRBs would
provide an important probe of the IGM at high-redshifts (Lamb \& Reichart
2000).

\acknowledgments{ 
A valuable contribution to
the early stages of this work was made by Pierre Demarque, who adapted
the Yale stellar structure code to compute models of very
massive zero-metallicity stars; although the results reported in this
paper were obtained with a more efficient code kindly supplied by
Dimitar Sasselov, the models of Pierre Demarque provided an important
check on their correctness. 
The authors wish to thank Joachim Puls for the assistance
in using his unified model atmosphere code, and
Michael Shull and Jason Tumlinson for
making available to us their composite Population III spectrum.
We are grateful to Martin Haehnelt and   
Martin Rees for helpful discussions. VB thanks
Paolo Coppi for
financial support under NASA grant NAG5-7074, and AL acknowledges
support by
NASA
grant NAG 5-7768 and NSF grant AST-0071019. }

%\vfill\eject

\clearpage

\begin{deluxetable}{llccl} 
\footnotesize
\tablewidth{13.cm}
\tablecaption{Production of Ionizing Photons $^{a}$ \label{tab1}}
\tablecolumns{5}
\tablehead{
\colhead{} &
\colhead{$E$} &
\colhead{$\dot{N}_{ion}$} &  
\colhead{$Y$} &
\colhead{$f_{\star}/\eta_{10}$}\\
\colhead{} &
\colhead{[eV]} &
\colhead{[photons s$^{-1}$ $M_{\odot}^{-1}$]} &  
\colhead{} &
\colhead{}
} 
\startdata
H I & $13.6$ & $1.6\times 10^{48}$ & $\sim 16$ & $1\times 10^{-4}$ \nl
He I & $24.6$ & $1.1\times 10^{48}$ & $\sim 14$ & $1\times 10^{-5}$ \nl
He II & $54.4$ & $3.8\times 10^{47}$ & $\sim 75$ & $4\times 10^{-5}$ \nl
\enddata
\tablenotetext{a}{The production rates are calculated for a $1000 M_{\odot}$
star. Within a factor of two, these rates apply to all stars with mass
above $300 M_{\odot}$.}
\tablecomments{$E$ is the ionization potential, $\dot{N}_{ion}$ is the
production rate of ionizing photons per unit stellar mass, $Y$ is the ratio of
$\dot{N}_{ion}$ for the cases of a heavy IMF and a Salpeter IMF, and
$f_{\star}$ is the required baryon fraction that needs
to be converted into stars in order to produce 
$10\eta_{10}$ ionizing photons per species particle (H I, He I, or He II).}
\end{deluxetable}

\clearpage

% fig.1

\setcounter{figure}{0}
\thispagestyle{empty}

\begin{center} % fig.1
\epsfig{file=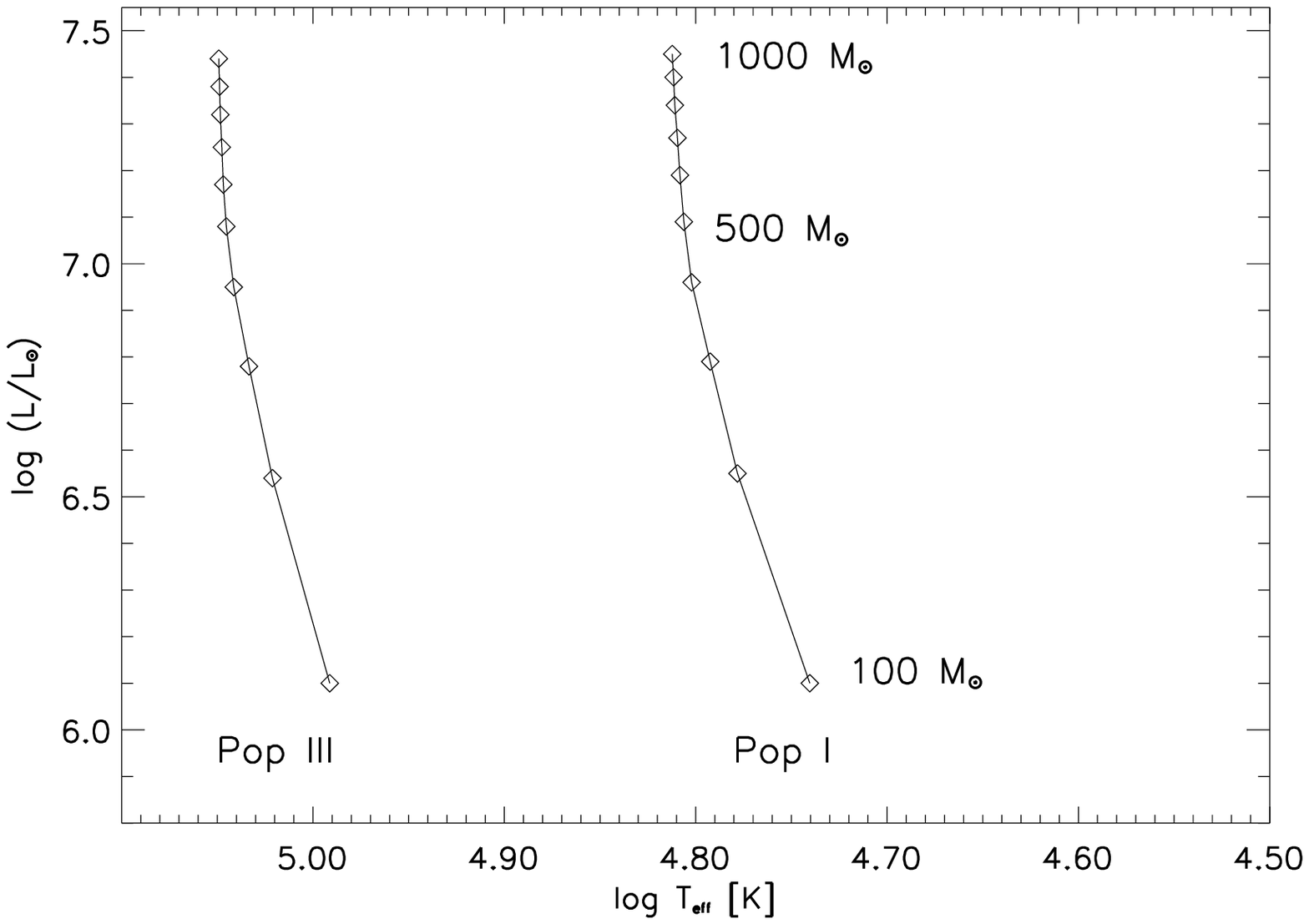,width=16.cm,height=12.6cm}
%%%%\vspace{10pt}
\figcaption{Zero-age main sequence of very massive stars.  {\it Left solid
line:} Population III zero-age main sequence (ZAMS).  {\it Right solid
line:} Population I ZAMS.  In each case, stellar luminosity (in
$L_{\odot}$) is plotted vs.  effective temperature (in K).  {\it
Diamond-shaped symbols:} Stellar masses along the sequence, from $100
M_{\odot}$ (bottom) to $1000 M_{\odot}$ (top) in increments of $100
M_{\odot}$.  The Population III ZAMS is systematically shifted to higher
effective temperature, with a value of $\sim 10^{5}$ K which is
approximately independent of mass. The luminosities, on the other hand, are
almost identical in the two cases.
\label{fig1}}
\end{center}
\clearpage

\thispagestyle{empty}

\begin{center} % fig.2
\epsfig{file=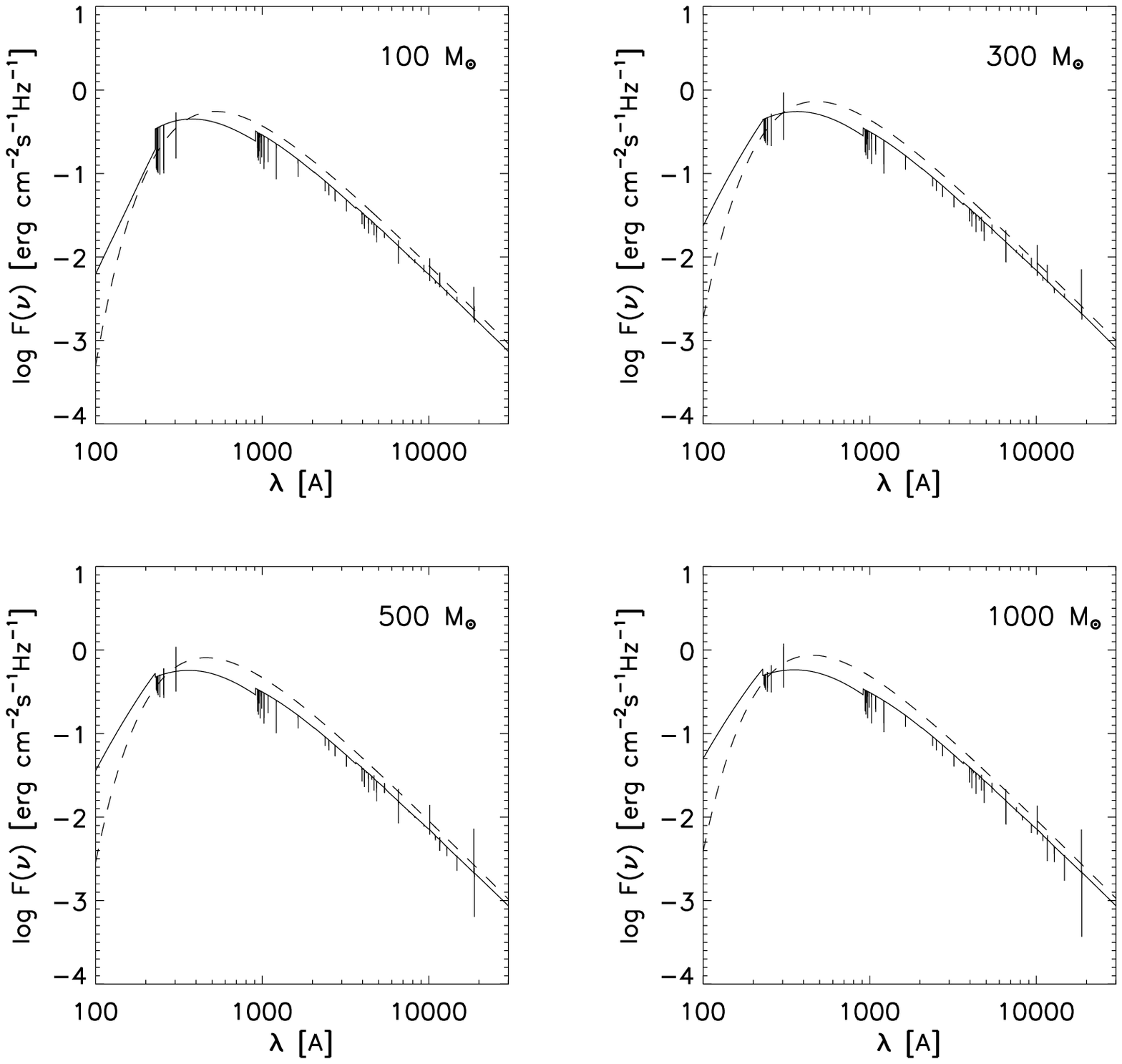,width=16.cm,height=12.6cm}
%%%%\vspace{10pt}
\figcaption{Spectra of very massive Population III stars.
{\it Solid lines:} Flux at the surface of the star (in units of
erg s$^{-1}$ cm$^{-2}$ Hz$^{-1}$) vs. wavelength (in \AA).
{\it Dashed lines:} Corresponding black-body spectrum, evaluated at the
respective effective temperatures.
Lines are due to transitions of H and He II in the EUV, UV, optical, and IR
bands.
The spectra for $M = 100, 300, 500, $ and $1000 M_{\odot}$ are remarkably
similar, differing only slightly in the strengths of the He II and H I 
ionization edges at $228 \mbox{\,\AA}$ and $912 \mbox{\,\AA}$, respectively.
\label{fig2}}
\end{center}
\clearpage

\thispagestyle{empty}

\begin{center} % fig.3
\epsfig{file=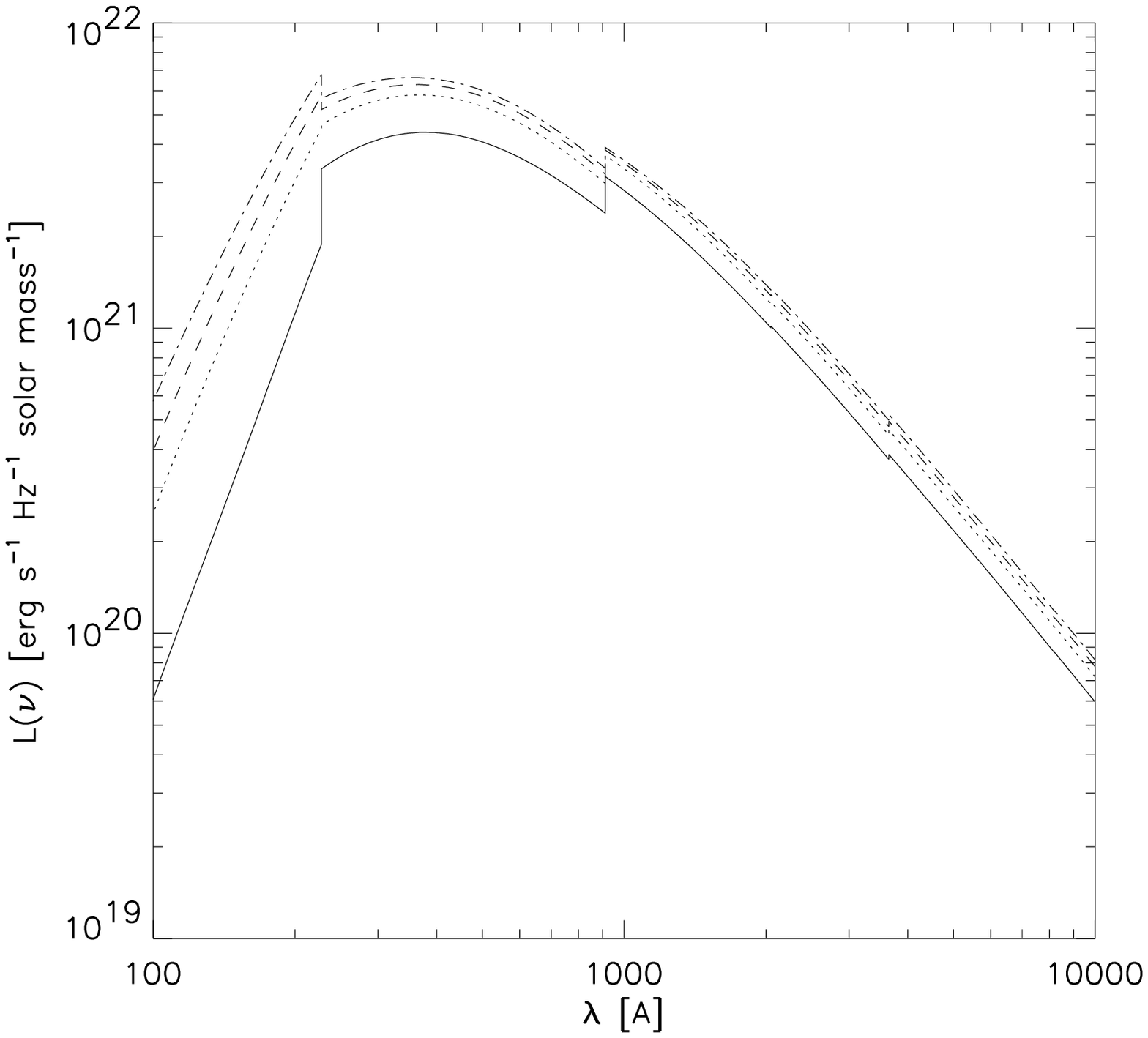,width=16.cm,height=12.6cm}
%%%%\vspace{10pt}
\figcaption{The normalized spectral energy distribution in the continuum.
Luminosity per unit stellar mass (in erg s$^{-1}$ Hz$^{-1}$ solar mass$^{-1}$) vs.
wavelength (in \AA).
{\it Solid line:} $100 M_{\odot}$.
{\it Dotted line:} $300 M_{\odot}$.
{\it Dashed line:} $500 M_{\odot}$.
{\it Dot-dashed line:} $1000 M_{\odot}$.
When plotted per unit mass, the spectra admit an almost universal form for $M>300M_{\odot}$.
\label{fig3}}
\end{center}
\clearpage

\thispagestyle{empty}

\begin{center} % fig.4
\epsfig{file=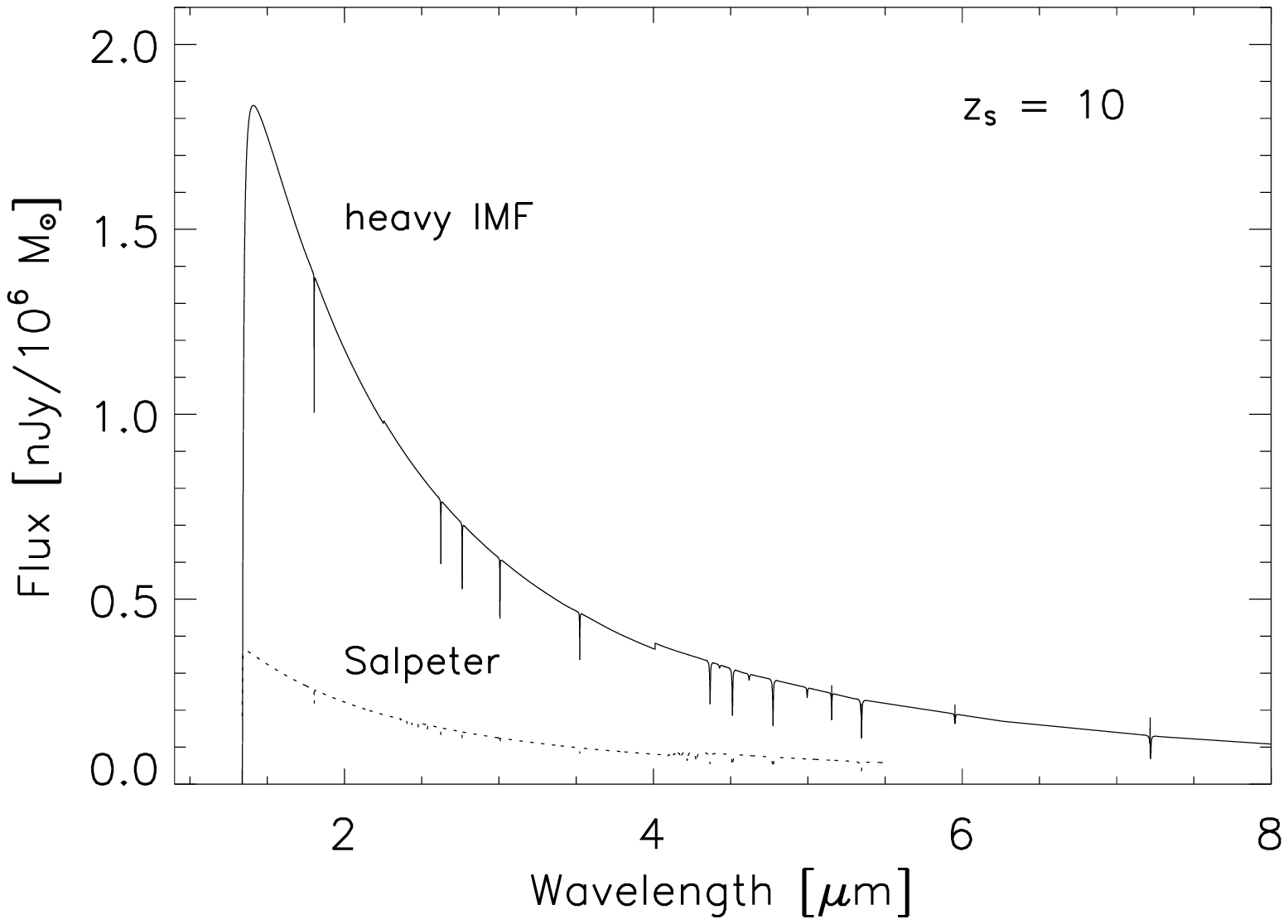,width=16.cm,height=12.6cm}
%%%\vspace{10pt}
\figcaption{The predicted flux from a Population III star cluster at
$z_{s}=10$.  Observed flux (in $\mbox{nJy}$ per $10^{6}M_{\odot}$ of stars)
vs. wavelength (in $\mu$m).  We assume a flat universe with
$\Omega_{\Lambda}=0.7$, and a Hubble constant of $H_{0}=65$ km s$^{-1}$
Mpc$^{-1}$.  {\it Solid line:} The case of a heavy IMF.  {\it Dotted line:}
The comparison case of a standard Salpeter IMF, where the composite
spectrum is taken from Tumlinson \& Shull (2000).  The cutoff below
$\lambda_{obs} = 1216\mbox{\,\AA \,} (1+z_{s})=1.34\mu$m is due to complete
Gunn-Peterson absorption.  It can be seen that for the same amount of total
stellar mass, the observable flux is larger by an order of magnitude for
stars which are distributed according to a heavy IMF.
\label{fig4}}
\end{center}

\end{document}